\documentclass[aps,pra,twocolumn,showpacs,groupedaddress,superscriptaddress,nofootinbib]{revtex4-1}
\usepackage{bm,graphicx,amsmath}
\usepackage[utf8]{inputenc}
\usepackage[T1]{fontenc}
\usepackage{lmodern}
\usepackage{amssymb}
\usepackage{amsmath}
\usepackage{epsfig}
\usepackage{natbib}
\usepackage{color}
\usepackage{bm}
\usepackage[pdftex,colorlinks,linkcolor=blue,citecolor=blue,urlcolor=blue]{hyperref}
\usepackage{array}
\usepackage{booktabs}
\usepackage[normalem]{ulem}
\usepackage{color}

\newcommand{\comment}[1]{}

\definecolor{gray}{gray}{0.6}

\begin{document}
\title{Geometrically induced complex tunnelings for ultracold atoms carrying orbital angular momentum}
\author{J. Polo}
\affiliation{Departament de F\'{\i}sica, Universitat Aut\`{o}noma de Barcelona, E-08193 Bellaterra, Spain} 
\author{J. Mompart}
\affiliation{Departament de F\'{\i}sica, Universitat Aut\`{o}noma de Barcelona, E-08193 Bellaterra, Spain} 
\author{V. Ahufinger}
\affiliation{Departament de F\'{\i}sica, Universitat Aut\`{o}noma de Barcelona, E-08193 Bellaterra, Spain} 
\email[]{Juan.Polo@uab.cat}
\date{\today}
\begin{abstract}
We investigate the dynamics of angular momentum states for a single ultracold atom trapped in two dimensional systems of sided coupled ring potentials. The symmetries of the system show that tunneling amplitudes between different ring states with variation of the winding number are complex. In particular, we demonstrate that in a triangular ring configuration the complex nature of the cross-couplings can be used to geometrically engineer spatial dark states to manipulate the transport of orbital angular momentum states via quantum interference.
\end{abstract}


\maketitle
\section{INTRODUCTION}
Tunneling is one of the paradigms of quantum mechanics and, recently, its control in the context of ultracold neutral atoms has been an issue of intense research. Pioneering experiments demonstrated dynamical tunneling suppression for a single-particle in a strongly driven double-well potential \cite{Oberthaler_2008} and for a Bose-Einstein condensate (BEC) in a strongly driven optical lattice \cite{Arimondo_2007}. The dynamical modification of tunneling rates allowed experimentally realizing \cite{Arimondo_2009} the driving-induced superfluid-Mott insulator transition \cite{Holthaus_2005,*Monteiro_2006} and, by independently tuning the coupling rates in different directions of a triangular lattice, simulating a large variety of magnetic phases and different types of phase transitions \cite{Sengstock_2011}.

The generation of artificial vector gauge potentials for ultracold atoms in one dimensional (1D) optical lattices has been demonstrated by inducing controllable complex tunneling amplitudes either by a suitable forcing of the optical lattice \cite{Windpassinger_2012} or by a combination of radio frequency and optical Raman coupling fields \cite{Spielman_2012}. In two dimensional (2D) optical lattices, the engineering of complex tunnelings has lead to the generation of staggered fluxes \cite{Aidelsburger_2011,*Mathey_2013}, the implementation of the Hofstadter Hamiltonian and the observation of large homogeneous artificial magnetic fields \cite{Bloch_2013}, as well as the realization of the topological Haldane model \cite{Esslinger_2014} and of the Harper and Weyl Hamiltonians \cite{Ketterle_2013,*Buljan_2015}.

In this article we demonstrate that complex tunneling amplitudes appear naturally in the dynamics of orbital angular momentum states for a single ultracold atom trapped in 2D systems of sided coupled cylindrically symmetric identical traps.
We focus on ring shaped traps, which are currently implemented for ultracold atoms by means of the optical dipole force or magnetic trapping.  Techniques for the first case include optically plugging magnetic traps \cite{Naik_2005,*Phillips_2007}, the use of static Laguerre-Gauss beams \cite{Wright_2000,*Moulder_2012,*Moulder_2013,*Courtade_2006,*Olson_2007}, painting potentials \cite{Schnelle_2008,*Houston_2008,*Henderson_2009}, time averaged potentials \cite{Arnold_2007,*Arnold_2012} and conical refraction \cite{Turpin_2015}. Alternatively, magnetic rings traps can be implemented using static magnetic fields \cite{Sauer_2001,*Wu_2004,*Arnold_2006,*Weiss_2015}, time-averaged magnetic fields \cite{Arnold_2004,*Gupta_2005,*West_2012}, by induction \cite{Griffin_2008,*Pritchard_2012} and using radiofrequency adiabatic potentials \cite{Sherlock_2011}.
Ring traps for ultracold atoms are one of the simplest geometries that lead to non-trivial loop circuits in the emerging field of atomtronics \cite{Holland_2007}, which explores the use of neutral atoms to build analogues of electronic circuits and devices. Specifically, BECs in a ring trap with one \cite{Campbell_2011,*Campbell_2013,*Campbell_2014,*Eckel_2014} or two \cite{Boshier_2013,Campbell_2014_1,*Ryu_2014} weak links have been shown to resemble the physical behaviour of superconducting quantum interference devices (SQUIDs). Atomic SQUIDs in a ring lattice have also been proposed \cite{Kwek_2014,*Amico_2015}. 

Specifically, we consider two 2D in-line ring potentials and three 2D rings in a triangular configuration. The full dynamics
Hilbert space consists of a set of decoupled manifolds spanned by ring states with identical vibrational and orbital angular momentum quantum numbers. Recalling basic geometric symmetries of the system, we show that the tunneling amplitudes between different ring states, named {\it cross-couplings}, with (without) variation of the winding number, are complex (real). Moreover, we show that a complex {\it self-coupling} between states with opposite winding number within a ring arises due to the breaking of cylindrical symmetry induced by the presence of additional rings and that these complex couplings can be controlled geometrically. Although for two in-line rings, the complex cross-coupling contribution is shown to give a non-physically relevant phase, we demonstrate that, in a triangular ring configuration, it leads to the possibility of engineering spatial dark states, which allows manipulating the transport of angular momentum states via quantum interference. This triangular trapping configuration may open a myriad of possibilities when assumed to be the unit cell of a 2D lattice.

\section{TWO IN-LINE RING POTENTIALS}
\begin{figure}[h!]
\includegraphics[width=0.35\textwidth]{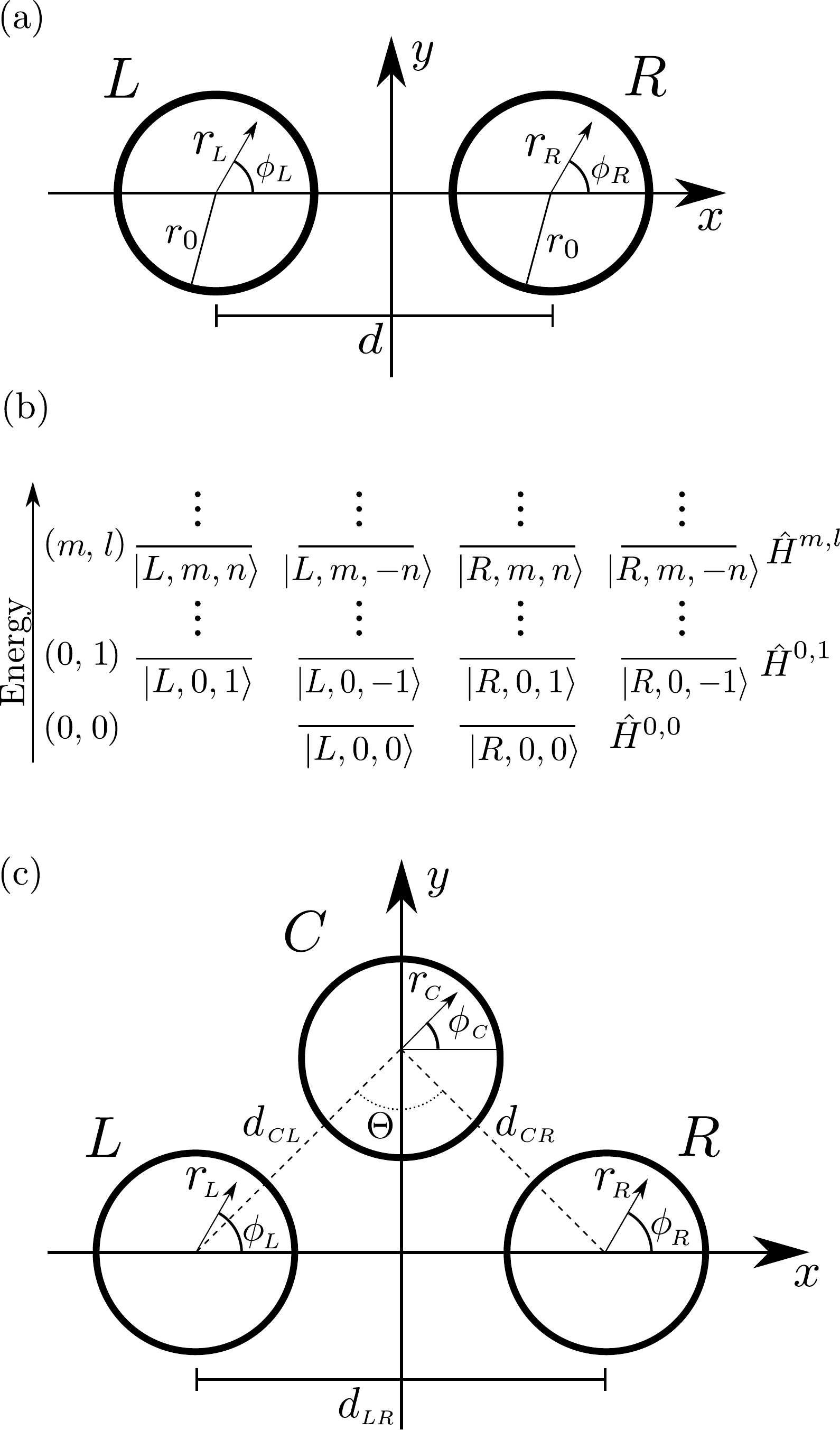}
\caption{(a) Two in-line ring potentials. (b) Sketch of the energy spectrum of the angular momentum eigenstates $|j,m,n\rangle$ for a single atom trapped in each of the two in-line rings, where $j=L,\;R$ indicates the ring, $m$ the transverse vibrational state, $n=\pm l$ the winding quantum number, and $l$ the orbital angular momentum. The Hamiltonian corresponding to the $(m,l)$ manifold is indicated by $\hat{H}^{m,l}$. (c) Three ring potentials in an isosceles triangular configuration.}
\label{fig:tworings}
\end{figure}

We consider a single atom trapped in a 2D system consisting of two in-line ring potentials of radius $r_0$ separated by a distance $d$, see Fig.~\ref{fig:tworings}(a). We define the radial coordinate $r_j$ and the azimuthal angle $\phi_j$ with respect to the center of each ring, where $j=L,\;R$ accounts for the left and right potentials, respectively.
The angular momentum eigenstates of each individual ring potential read:
\begin{equation}
\Psi_{j,m}^{n}(r_j,\phi_j)=\langle\vec{r}|j,m,n\rangle=\frac{1}{\sqrt{N}}\psi_{m}(r_j)e^{in(\phi_j-\phi_0)},\label{eq:states}
\end{equation}
where $j=L,\;R$; $n=\pm l$ is the winding number with $l\in \mathbb{N}^0$ being the orbital angular momentum quantum number, $\psi_{m}(r_j)$ with $m\in \mathbb{N}^0$ is the radial part of the wave function for the $m$ transverse vibrational state, $\phi_0$ is a free phase parameter, defined with respect to the $x$ axis, which sets the azimuthal phase origin, and $N$ is a normalization constant.

Without loss of generality, we assume $\psi_m (r_j)$ to be real and, thus, the phases of the, in general, complex tunneling amplitudes will be only determined by $\phi_0$. To obtain these phases, we recall that two identical in-line rings present two symmetries, the $x$ and the $y$ mirrors defined as:
\begin{subequations}
\begin{align}
&\hat{M}_{x}:\quad
(x,\; y) \longrightarrow (x,\; -y),\\
&\hat{M}_{y}:\quad
(x,\; y)\longrightarrow (-x,\; y),
\end{align}
\end{subequations}
respectively. The effects of such transformations on the angular momentum eigenstates, Eq.~(\ref{eq:states}), are:
\begin{subequations}
\label{eq:symmetries}
\begin{gather} 
\hat{M}_{x}|j,m,n\rangle=e^{-2in\phi_0}|j,m,-n\rangle,\\
\hat{M}_{y}|j,m,n\rangle=e^{-2in\phi_0}e^{in\pi}|k,m,-n\rangle\text{ for $j\neq k$}.
\end{gather}
\end{subequations}

Assuming $\sigma_m \ll r_0 \ll d$, where $\sigma_m$ is the radial width of the atom in the $m$ vibrational state of any of the two rings, the total Hamiltonian of the system reads:
\begin{equation}
\hat{H}_{T}=\sum_{m\geq 0}\hat{H}^{m,0}+\sum_{m\geq 0} \sum_{l>0}\hat{H}^{m,l},
\end{equation}
where $\hat{H}^{m,0}$ accounts for the two-state Hamiltonian associated to $|L,m,0\rangle$ and $|R,m,0\rangle$, and
$\hat{H}^{m,l}$ corresponds to the four-state Hamiltonian (FSH) for the $(m,l)$ combination with $l\neq0$, whose basis is formed by four degenerate angular momentum eigenstates $|L,m,\pm n\rangle$, $|R,m,\pm n\rangle$, see Fig.~\ref{fig:tworings}(b):
\begin{gather}
\hat{H}^{m,l}=\frac{\hbar}{2}\sum_{j,k=L,R}\sum_{n=\pm l}\left(J_{j,n}^{k,n}(m)|j,m,n\rangle\langle k,m,n|+\right.\nonumber \\ \left. J_{j,n}^{k,-n}(m)|j,m,n\rangle\langle k,m,-n|\right),\label{eq:FSH}
\end{gather}
where $(\hbar/2)J_{j,n}^{j,n}(m)$ are the eigenenergies of a single atom in an uncoupled ring, $J_{j,n}^{j,-n}(m)$ are the self-coupling tunnelings, and $J_{j,n}^{k,-n}(m)$ and $J_{j,n}^{k,n}(m)$ with $j\neq k$ are the cross-coupling tunnelings.\\\indent
The Hamiltonian describing this system is invariant under $\hat{M}_x$ and $\hat{M}_y$ transformations and, consequently, under parity $\hat{P}=\hat{M}_x\hat{M}_y$. Using the symmetry transformations acting on the angular momentum states, Eqs.~(\ref{eq:symmetries}), we obtain that:
\begin{subequations}
\label{eq:tworingssymmetries}
\begin{align}
\hat{M}_{x}:\qquad\qquad\qquad J_{j,n}^{k,n}&=J_{j,-n}^{k,-n},\\
e^{-2in\phi_0}J_{j,n}^{k,-n}&=e^{2in\phi_0}J_{j,-n}^{k,+n},\\
\hat{M}_{y}:\qquad J_{j,n}^{k,n}=J_{k,-n}^{j,-n},&\quad J_{j,n}^{j,n}=J_{k,-n}^{k,-n}\text{ for $j\neq k$},\\
e^{-2in\phi_0}J_{j,n}^{k,-n}&=e^{2in\phi_0}J_{k,-n}^{j,n}\text{ for $j\neq k$},\\
e^{-2in\phi_0}J_{j,n}^{j,-n}&=e^{2in\phi_0}J_{k,-n}^{k,n}\text{ for $j\neq k$},
\end{align}
\end{subequations}
These relations between the couplings along with the Hermiticity of the Hamiltonian, reduce the parameter space to only three different couplings: a real coupling $J_{L,n}^{R,n}$ and two complex ones $J_{L,n}^{L,-n}=|J_{L,n}^{L,-n}|e^{2in\phi_0}$ and $J_{L,n}^{R,-n}=|J_{L,n}^{R,-n}|e^{2in\phi_0}$.
For two rings, we can fix $\phi_0$ to any arbitrary value. Thus, for $\phi_0=0$ all couplings become real and the four-state Hamiltonian $\hat{H}^{m,l}$ reads:
\begin{equation}
\label{eq:HFS2}
\hat{H}^{m,l}=\frac{\hbar}{2}
\begin{pmatrix}
	0		&	J_{L,n}^{L,-n}	&	J_{L,n}^{R,n}	&	J_{L,n}^{R,-n}\\
J_{L,n}^{L,-n}	&		0		&	J_{L,n}^{R,-n}  &	J_{L,n}^{R,n}\\
J_{L,n}^{R,n}	&	J_{L,n}^{R,-n}	&		0		&	J_{L,n}^{L,-n}\\
J_{L,n}^{R,-n}	&	J_{L,n}^{R,n}	&	J_{L,n}^{L,-n}	&		0	
\end{pmatrix},
\end{equation}%
where we have subtracted the common energy from the diagonal. In fact, the complex nature of the self-couplings and cross-couplings with winding number exchange does not play any physical role in the two in-line ring configuration. However, as detailed below, it will become crucial when studying the dynamics of more than two coupled rings.
Note also that, although for a single ring $J_{j,n}^{j,-n}=0$, in the case of two in-line coupled rings a non null coupling appears between opposite winding number states in the same ring. This coupling emerges due to the breaking of cylindrical symmetry in the system \cite{Pethick_2007}, produced by the presence of the second ring.
 
To numerically investigate the free dynamics of a single atom of mass $M$ in two in-line rings, Fig.~\ref{fig:tworings}(a), we consider two radially truncated harmonic ring potentials of frequency $\omega$. The atom is initially trapped in state $|L,0,1\rangle$.
The distance between rings is kept fixed during the dynamics at $d=14$ and $r_0=5$, all in 1D radial harmonic oscillator (h.o.) units.

\begin{figure}
\includegraphics[width=0.5\textwidth]{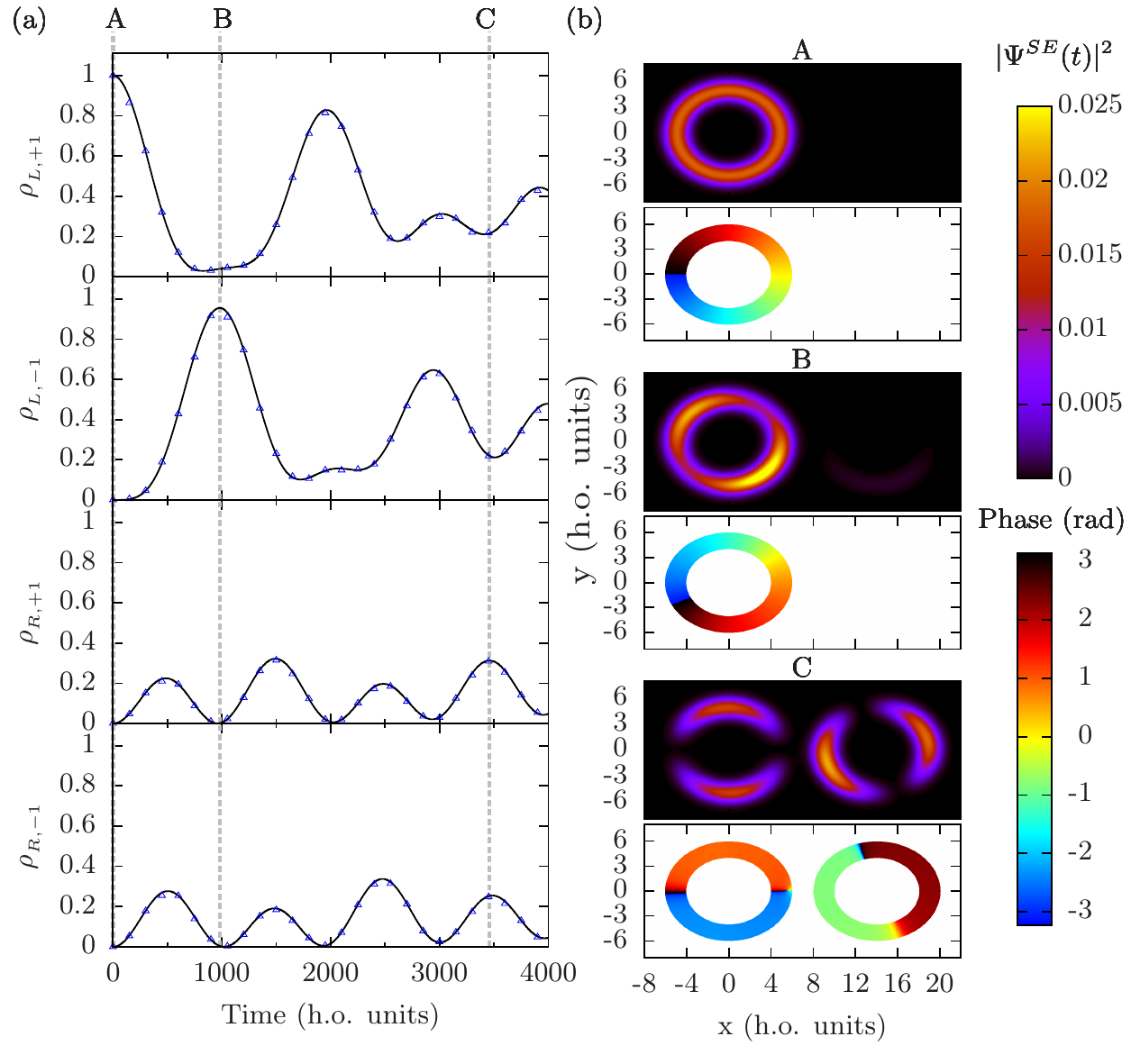}
\caption{(a) Temporal evolution of the population of each angular momentum state involved in the dynamics, $\rho_{j,\pm1}=|\langle\Psi(t)|j,0,\pm1\rangle|^2$, where $j=L,\;R$, using the numerically integrated 2D SE (points) and the FSH, Eq.~(\ref{eq:HFS2}), (lines). (b) Atomic probability density (upper plots) and phase distribution (lower plots) of the state of the system at times $A$, $B$, $C$ in Fig.~\ref{fig:FSM}(a). The phase is only plotted where the probability density is non negligible. For the parameters see text.}
\label{fig:FSM}
\end{figure}
Fig.~\ref{fig:FSM}(a) shows the temporal evolution of the populations of the four angular momentum states of the $(0,1)$ manifold by using the FSH, Eq.~(\ref{eq:HFS2}), and the numerical integration of the full 2D Schr\"odinger equation (SE). The perfect agreement has been achieved by taking $J_{L,1}^{R,1}=3.06\times10^{-3}$, $J_{L,1}^{R,-1}=3.28\times10^{-3}$ and $J_{L,1}^{L,-1}=-4.06\times10^{-4}$ in the FSH, in h.o. units. These values of the tunneling amplitudes for the FSH have been obtained by numerically constructing the eigenstates basis of the total system. Then, by taking states with angular momentum $n=\pm1$, that are degenerate in the basis $|j,0,\pm1\rangle$ (with $j=L,\,R$), we are able to build a FSH whose diagonalization gives rise to a simple relation between the eigenenergies of the total system and the tunneling amplitudes of the mentioned FSH.

The dynamics shows that the population is being initially transferred from $|L,0,1\rangle$ to states $|R,0,\pm1\rangle$ to come back again ($B$ in Fig.~\ref{fig:FSM}(a)) to the left trap but mostly with opposite winding number, i.e., to state $|L,0,-1\rangle$. From the FSH numerical simulations, we have checked that there are no complete population oscillations between states $|L,0,1\rangle$ and $|L,0,-1\rangle$ due to the self-coupling contribution $J_{L,1}^{L,-1}$. Note that since the self-coupling appears due the asymmetry of the system and the cross-coupling has two contributions, the asymmetry and the tunneling through the kinetic energy term between the two rings, the tunneling amplitude of the self-coupling is in general smaller than the cross-coupling. In particular, in our example the self-coupling is roughly one order of magnitude smaller than the cross-couplings.

From the integration of the 2D SE, Fig.~\ref{fig:FSM}(b) shows the atomic probability density and the phase distribution at times $A$, $B$ and $C$ in Fig.~\ref{fig:FSM}(a). $A$ corresponds to the initial state $|L,0,1\rangle$. In $B$, we observe the appearance of two minima in the left ring probability density produced by the coexistence of states $|L,0,1\rangle$ and $|L,0,-1\rangle$. Finally, $C$ corresponds to the state formed by an approximately equally weighted superposition of the four states of the $(0,1)$ manifold. Accordingly, two density nodes appear in each ring. 

\section{TRIANGULAR CONFIGURATION}

We consider now three identical ring potentials (labeled $L$, $C$, $R$ from left, central and right) of radius $r_0$ in a triangular configuration with distances between their centers $d_{CL}=d_{CR}\equiv d$ and $d_{LR}=2d\sin(\Theta/2)$, see Fig.~\ref{fig:tworings}(c).
Considering the angular momentum eigenstates of each ring, Eq.~(\ref{eq:states}), the bare energy spectrum of a single atom trapped in any of the three ring potentials is formed by a set of manifolds of six degenerate angular momentum states, for each $(m,l)$ combination with $l\neq0$, $|L,m,\pm n\rangle$, $|C,m,\pm n\rangle$ and $|R,m,\pm n\rangle$ plus manifolds of three degenerate states of null orbital angular momentum $|L,m,0\rangle$, $|C,m,0\rangle$ and $|R,m,0\rangle$.
Following the procedure developed for the two in-line rings configuration, the total Hamiltonian of the system can be written as a direct sum of the three-state Hamiltonians with $l=0$, plus six-state Hamiltonians (SSH) for each $(m,l)$ combination with $l\neq0$.

By assuming that the rings $L$ and $R$ are decoupled, i.e., $d_{LR}\gg d$, we can describe the system as two sets of two in-line coupled rings ($C$-$L$ and $C$-$R$). By setting the free phase parameter $\phi_0=0$ with respect to the $C$-$L$ axis, we can use Eqs.~(\ref{eq:tworingssymmetries}) with $\phi_0=0$ and $j,k=L,C$ to determine the $C$-$L$ couplings, which will be real. Similarly, the relations between the $C$-$R$ couplings can be obtained using Eqs.~(\ref{eq:tworingssymmetries}) with $\phi_0=\Theta$ and $j,k=R,C$. Thus, by means of the geometrical parameter $\Theta$ one can manipulate the phases of the complex tunnelings. Using the $\hat{M}_y$ symmetry of the three triangular configuration we find an additional relation:
\begin{equation}
\label{eq:SSMcouplings}
e^{-in\Theta}J_{L,n}^{L,-n}=e^{in\Theta}J_{R,-n}^{R,n},
\end{equation}
which relates the two sets of two in-line systems $C$-$L$ and $C$-$R$. Considering the full system $L$-$C$-$R$, one can see that the central self-coupling is in fact a combination of the ones obtained in the two sets of systems, $C$-$L$ and $C$-$R$, and thus, it becomes $J_{L,+n}^{L,-n}+J_{R,+n}^{R,-n}=J_{L,+n}^{L,-n}(1+e^{2in\Theta})$.
Note that $\Theta=\pi$ in Fig.~\ref{fig:tworings}(c) corresponds to three in-line ring potentials for which $J_{C,n}^{C,-n}=2|J_{L,n}^{L,-n}|$, while $\Theta=\pi/2$ corresponds to an isosceles triangle configuration for which $d_{LR}=\sqrt{2}d$, and $J_{C,n}^{C,-n}=0$.

\begin{figure}
\includegraphics[width=0.5\textwidth]{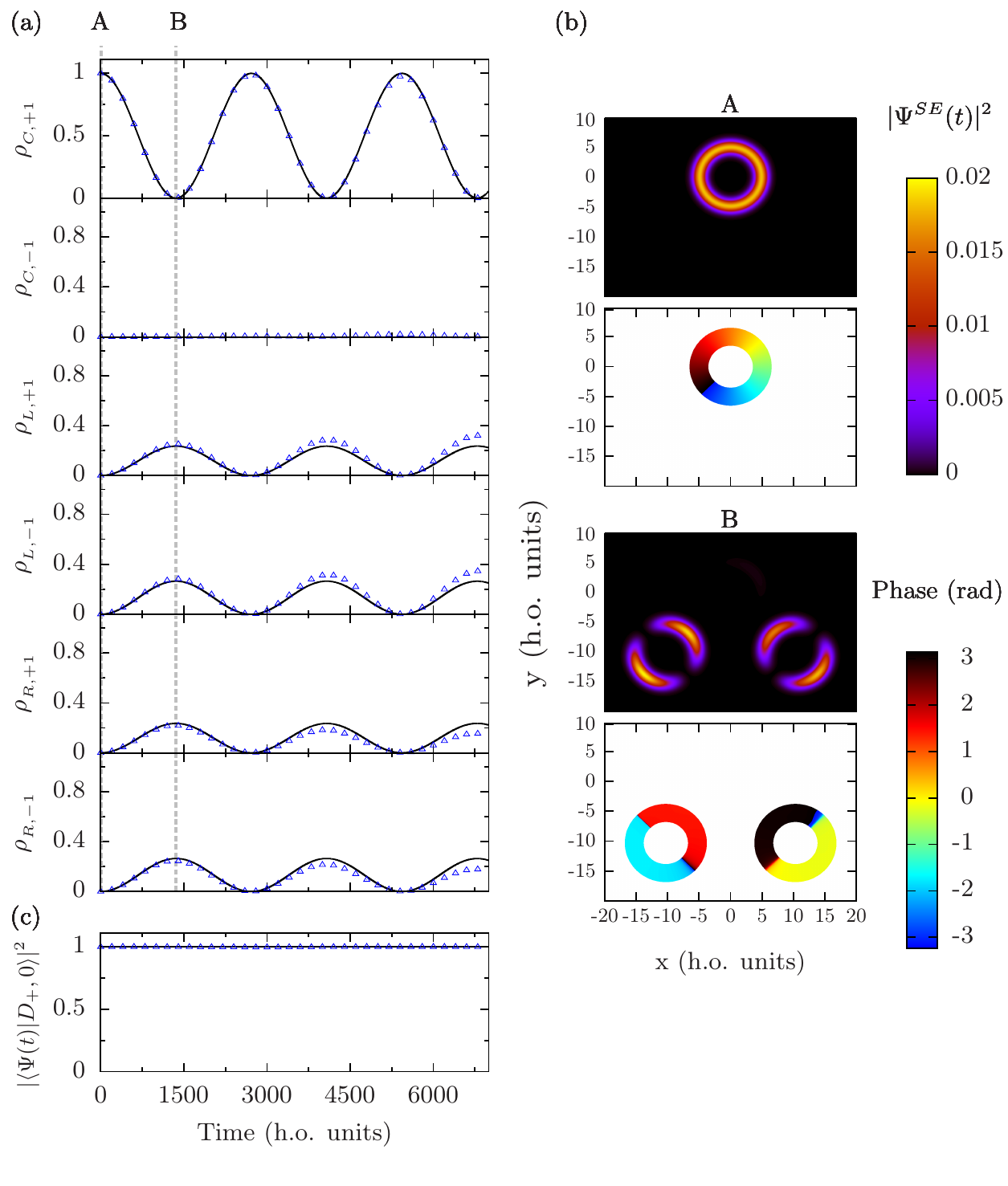}
\caption{(a) Temporal evolution of the population of each angular momentum state involved in the dynamics, $\rho_{j,\pm1}=|\langle\Psi(t)|j,0,\pm1\rangle|^2$, where $j=C,\;L,\;R$, using the numerically integrated 2D SE (points) and the SSH (lines) when the initial state of the system is $|C,0,1\rangle$. (b) Probability density (upper plots) and phase distribution (lower plots) of the state of the system at times $A$ and $B$ in Fig.~\ref{fig:SSM}(a). (c) Temporal evolution of the population of the dark state $|D_{+},0\rangle$ using the numerically integrated SE (points) and the SSH (lines) when the system is initialized in this state. Phase is only plotted where the probability density is non negligible. For the parameters see text.}
\label{fig:SSM}
\end{figure}

Assuming that the self-couplings are negligible compared to the cross-couplings and taking $\Theta=(2s+1)\pi /2n$ with $s\in \mathbb{N}$, each six-state manifold, for a given $(m,l)$ combination, can be mapped into two two-level systems $|C,m,n\rangle\leftrightarrow|B_{+},m\rangle$ and $|C,m,-n\rangle\leftrightarrow|B_{-},m\rangle$ with the two spatial bright states defined as:
\begin{equation} 
\label{eq:brightstates}
|B_{\pm},m\rangle\equiv\frac{1}{J}\left(J_{L,n}^{C,n}|S,m,\pm n\rangle+J_{L,n}^{C,-n}|A,m,\mp n\rangle\right),
\end{equation} 
plus two spatial dark states decoupled from the dynamics:
\begin{equation} 
\label{eq:darkstates}
|D_{\pm},m\rangle\equiv\frac{1}{J}\left(J_{L,n}^{C,-n}|S,m,\pm n\rangle-J_{L,n}^{C,n}|A,m,\mp n\rangle\right),
\end{equation} 
where:
\begin{subequations}
\begin{align} 
|S,m,\pm n\rangle&=\frac{1}{\sqrt{2}}\left(|L,m,\pm n\rangle+|R,m,\pm n\rangle\right)\\
|A,m,\pm n\rangle&=\frac{1}{\sqrt{2}}\left(|L,m,\pm n\rangle-|R,m,\pm n\rangle\right),
\end{align} 
\end{subequations}
and $J=\sqrt{|J_{L,n}^{C,n}|^2+|J_{L,n}^{C,-n}|^2}$. From these definitions, it is straightforward to check that: $\langle C,m,n |\hat{H}^{m,l}|D_{+},m\rangle=\langle C,m,-n |\hat{H}^{m,l}|D_{-},m\rangle=0$ and that the only remaining couplings are: $\langle C,m,n |\hat{H}^{m,l}|B_{+},m\rangle=\langle C,m,-n |\hat{H}^{m,l}|B_{-},m\rangle=\sqrt{2}J$.

To numerically study the free dynamics of a single atom trapped in a three ring configuration, Fig.~\ref{fig:tworings}(c), we consider three harmonic ring potentials of frequency $\omega$ centered at the vertices of an isosceles triangle with $\Theta=\pi/2$.
In Fig.~\ref{fig:SSM}(a) we plot the temporal evolution of the populations of the angular momentum states $|j,0,\pm1\rangle$, with $j=L,\;C,\;R$, using the SSH and the numerical integration of the full 2D SE, with the atom being initially in $|C,0,1\rangle$. We observe, as predicted above, that the population oscillates between states $|C,0,1\rangle$ and $|B_{+},0\rangle$. The values of the couplings in the SSH are $J_{L,1}^{C,1}=1.12\times10^{-3}$, $J_{L,1}^{C,-1}=1.18\times10^{-3}$ and $J_{L,1}^{L,-1}=-7.76\times10^{-5}$ and we fix $r_0=5$ and $d=14.5$, in h.o. units. Note that the ratio between the self-coupling of the lateral rings and the cross-coupling tunneling amplitudes in this triangular configuration is smaller than for the two-ring case. This simply occurs since we are considering a larger distance and the exponential decay of the self-coupling is faster than for the cross-coupling. Moreover, as discussed along the lines following Eq.~(8), note that for this particular triangular geometry the self-coupling contribution of the middle ring is completely suppressed.

Fig.~\ref{fig:SSM}(b) shows the density and phase snapshots at times $A$ and $B$ in Fig.~\ref{fig:SSM}(a). $A$ corresponds to the initial state $|C,0,1\rangle$. In $B$ we can see that the initial state has been fully transferred to the bright state $|B_{+},0\rangle$, as given in Eq.~(\ref{eq:brightstates}), which corresponds to an almost equally weighted combination of the four states $|L,0,\pm1\rangle$ and $|R,0,\pm1\rangle$. In Fig.~\ref{fig:SSM}(c), we demonstrate the existence of spatial dark states, Eq.~(\ref{eq:darkstates}), by using both the SSH and the numerically integrated 2D SE. Specifically, we select as initial state $|D_{+},0\rangle$ and let the system evolve freely. We observe that the dark state remains decoupled from the dynamics and, therefore, states $|C,0,\pm1\rangle$ are never populated.

\section{CONCLUSIONS}
 
We have studied the dynamics of the angular momentum states of a single ultracold atom trapped in 2D systems of sided coupled identical ring traps.
We have demonstrated that the couplings between states of different rings with different winding number are complex and that the breaking of the cylindrical symmetry induced by the presence of the neighboring rings produces a complex self-coupling between angular momentum states with opposite winding number within the same ring.
Worth to highlight, the results here derived are solely based in the mirror symmetries that exhibit sided coupled cylindrically symmetric identical potentials carrying angular momentum atomic states. Thus, they could be applied not only to rings but also, for instance, to 2D identical isotropic harmonic traps. On the other hand, although the article has been focused on the single atom case, it would be interesting to extend our results to BECs trapped in ring potentials to investigate the role of the non-linearity in the self and cross-coupling tunneling amplitudes. Note that, even though most of current experimental setups consider ring radii larger than those discussed in the article, one of the main experimental short-term goals in atomtronics is to build-up smaller rings. For instance, rings with radii of $4\,\mu$m were used to build and investigate a SQUID in \cite{Boshier_2013}.
The ring radius of our examples shown here would correspond to approximately $5.6\,\mu$m when using the same radial trapping frequency as in \cite{Boshier_2013}.

In a triangular ring configuration, we have demonstrated that the complex nature of the couplings between angular momentum states yields spatial bright and dark states that depend on the system's geometry. Thus, these complex couplings could be used in 2D trapping configurations, e.g., 2D optical lattices, of cylindrically symmetric identical traps to manipulate the dynamics of ultracold atoms by means of the constructive (destructive) quantum interference associated with spatial bright (dark) states. Note, finally, that the particular dynamical evolutions induced by the complex tunnelings may be inferred through density measurements in current experimental setups \cite{Klitzing_2007_1,Boshier_2009,Kwek_2014,Campbell_2014}.

The authors acknowledge financial support through the Spanish and Catalan contracts FIS2014-57460-P and SGR2014-1639. J. Polo also acknowledges financial support from FPI Grant No. BES-2012-053447.
\bibliographystyle{apsrev4-1}

\bibliography{references}
\clearpage
\end{document}